\documentclass[12pt,preprint]{aastex}

\usepackage{emulateapj5}

\slugcomment{to Appear in the Astrophysical Journal, Letters}

\shorttitle{Recurrent Novae CI Aquilae}
\shortauthors{Hachisu and Kato}

\begin{document}

\title{PREDICTION OF SUPERSOFT X-RAY PHASE, HELIUM ENRICHMENT, AND
TURN-OFF TIME IN RECURRENT NOVA CI AQUILAE 2000 OUTBURST}


\author{Izumi Hachisu}
\affil{Department of Earth Science and Astronomy, 
College of Arts and Sciences, University of Tokyo,
Komaba, Meguro-ku, Tokyo 153-8902, Japan} 
\email{hachisu@chianti.c.u-tokyo.ac.jp}

\and

\author{Mariko Kato}
\affil{Department of Astronomy, Keio University, 
Hiyoshi, Kouhoku-ku, Yokohama 223-8521, Japan} 
\email{mariko@educ.cc.keio.ac.jp}

%




\begin{abstract}
     Recurrent nova CI Aquilae is still bright 300 days 
after the optical maximum, showing the slowest evolution
among recurrent novae.  We predict the turn-off time of CI Aql 
2000 outburst coming in August 2001 after a supersoft
X-ray source (SSS) phase lasts 250 days.
We also predict helium enrichment of ejecta, He/H$\sim 0.25$ by number.
Observational confirmations are urgently required.
Based on the optically thick wind mass-loss theory of
the thermonuclear runaway model, we have also estimated 
the WD mass to be $M_{\rm WD}= 1.2 \pm 0.05 ~M_\sun$ 
by fitting our theoretical light curves 
with the 1917 and 2000 outbursts.
The mass of the hydrogen-rich envelope on the WD 
is also estimated to be 
$\Delta M \sim 6 \times 10^{-6} M_\sun$ at the optical maximum, 
indicating an average mass accretion rate of
$\dot M_{\rm acc} \sim 0.7 \times 10^{-7} M_\sun$ yr$^{-1}$ 
during the quiescent phase between the 1917 and 2000 outbursts. 
Using these obtained values, we have consistently reproduced 
the light curve in quiescence as well as the two outburst phases.
Thus, we predict the turn-off time to be in August 2001 
for the 2000 outburst.
We strongly recommend soft X-ray observations to detect SSS
until August 2001 
because the massive wind phase have already ended 
in December 2000 followed by an SSS phase that very likely 
lasts until August 2001.
\end{abstract}




\section{INTRODUCTION}
     The second recorded outburst of CI Aquilae 
was discovered in April 2000 by \citet{tak00}
since the first recorded outburst in 1917 \citep{rei25,due87}. 
Thus, CI Aql has been recognized 
as a member of the recurrent novae \citep*{yam00}.
After optical brightness reached its maximum $V \sim 9$ 
in early May 2000, it rapidly decreased to 
$V \sim 13$ in about 50 days.  A plateau phase follows;
the brightness levelled off at $V \sim 13.5$. 
CI Aql is still as bright as $V \sim 14$ until now
about 300 days after maximum, being the longest record of
on-time among the recurrent novae (e.g., VSNET archives, 
{\tt http://www.kusastro.kyoto-u.ac.jp/vsnet/}).
\par
     Recurrent novae constitute a small class of cataclysmic 
variables, the outbursts of which repeat in a decade to a century.
Ejecta of the recurrent novae are not metal-enriched but similar
to the solar value ($Z\sim 0.02$), indicating that the WDs
in the recurrent novae are not eroded but have grown
in mass \citep*[e.g.,][]{sta85}.  
Masses of white dwarfs (WDs) in the recurrent novae are estimated to be 
so massive as the Chandrasekhar mass limit from the light
curve fitting \citep[e.g.,][]{hac99k,hkkm00,hac00ka,hac00kb,hac01k}.
These features strongly 
suggest that the recurrent novae are an immediate progenitor of
Type Ia supernovae (SNe Ia) if the WD consists of carbon and
oxygen \citep*{nom84}.  Evolutionary paths to SNe Ia via recurrent novae 
have been pointed out by \citet*{hkn99}.
\par 
     Recurrent novae are divided into three subclasses
depending on their orbital periods: T Pyx subclass with a dwarf
companion,  U Sco subclass with a slightly evolved main-sequence
(MS) or subgiant companion, and RS Oph subclass with 
a red giant companion \citep[e.g.,][]{war95}.
CI Aql has an orbital period of 0.618 days, belonging to the
U Sco subclass.   It is therefore likely that the light curve 
of CI Aql develops like U Sco.  
\par
     The 1999 outburst of U Sco was densely observed
from the rising phase to the final decay phase through 
the mid-plateau phase.  
The detection of supersoft X-rays during the plateau phase
\citep{kah99} strongly supports the theoretical description 
of nova outbursts
--- massive wind phase, steady hydrogen burning phase (no winds), and
final cooling phase, --- which corresponds to
the rapid decline, plateau, and final decline phases 
in the light curve \citep[e.g.,][]{hkkm00}, respectively.
Because the development of CI Aql outbursts is much slower than 
that of U Sco, there is still a chance to observe supersoft X-rays.
Supersoft X-rays are direct evidence of hydrogen shell burning 
on a WD so that they provide us valuable information on 
recurrent novae and SN Ia progenitors.
\par
     In this {\it Letter}, we determine the mass and the growth 
rate of the WD as accurately as possible by theoretically analyzing 
light curves.  Then,
(1) we predict the turn-off time of nuclear burning and 
the duration of supersoft X-ray source (SSS) phase of the present 
outburst and (2) examine the possibility whether or not
CI Aql will explode as an SN Ia.
In \S 2, we describe our model of CI Aql in quiescence and estimate
the inclination angle, the size and thickness of the accretion
disk, and so on.
In \S 3, the 1917 outburst is numerically analyzed and
we determine the mass of the WD.  
In \S4, the 2000 outburst is also numerically analyzed
and the duration of wind is estimated.  In \S 5, 
we predict the duration of the SSS phase.  Discussion follows in \S6.

\section{BASIC MODEL IN QUIESCENCE}
     CI Aql is an eclipsing binary.  
We have considered a binary model as illustrated 
in Figure \ref{ciaql_quiescence_config} which consists 
of (1) the WD photosphere, (2) the MS photosphere which fills
its Roche lobe, and (3) the accretion disk (ACDK). 
We assume that the size and thickness of the ACDK are given
by two parameters, $\alpha$ and $\beta$, as 
\begin{equation}
R_{\rm disk} = \alpha R_1^*,
\label{accretion-disk-size}
\end{equation}
and
\begin{equation}
h = \beta R_{\rm disk} \left({{\varpi} 
\over {R_{\rm disk}}} \right)^\nu,
\label{flaring-up-disk}
\end{equation}
where $R_{\rm disk}$ is the outer edge of the ACDK,
$R_1^*$ the effective radius of the inner critical Roche lobe 
for the WD component, and
$h$ the height of the surface from the equatorial plane,
$\varpi$ the distance on the equatorial plane 
from the center of the WD.  Here, we adopt $\varpi$-square law 
($\nu=2$) 
to mimic the effect of flaring-up at the rim of the ACDK
\citep*[e.g.,][]{sch97}.

\placefigure{ciaql_quiescence_config}

\par
     The accretion luminosity of the WD \citep*[e.g.,][]{sta88}, 
the viscous luminosity of the ACDK, 
and the irradiations by the WD \citep[e.g.,][]{sch97}
are also included.  
Here, we assume that the surfaces of the MS and of
the ACDK emit photons as a black-body at a local 
temperature of each surface patch area 
(see Fig. \ref{ciaql_quiescence_config}) 
which is heated by radiation from the WD.  
The efficiencies of irradiation of the MS and of the ACDK are assumed 
to be $\eta_{\rm MS}=0.5$ and $\eta_{\rm DK}=0.5$, respectively.
The numerical method is essentially the
same as that in \citet{hkkmn00} and fully described in \citet{hac01k}.
A circular orbit is assumed. 
Light curves of CI Aql in quiescence has been obtained by \citet{men95}. 
They determined the orbital period and ephemeris 
to be JD 2,448,412.167(25) + 0.618355$\times E$
at eclipse minima.
\par
     Figure \ref{vmag_quiescence} shows the averaged (smoothed)
observational points (open circles) together 
with our calculated $V$ light curves 
for the suggested WD mass of $M_{\rm WD}= 1.2 M_\odot$
($R_{\rm WD}= 0.0072 R_\odot$) and mass accretion rate of
$\dot M_{\rm acc} \sim 1 \times 10^{-7} M_\odot$ yr$^{-1}$
(see the following sections).
To fit our theoretical light curves with the
observational points, we have calculated $V$ light 
curves by changing the parameters of
$\alpha=0.5$---1.0 by 0.1 step, $\beta=0.05$---0.50 by 0.05 step,
$T_{\rm ph, MS}= 5000$---8000 K by 100 K step,
$T_{\rm ph, disk}= 3000$---6000 K by 500 K step,
and $i=70$---$90\arcdeg$ by $1\arcdeg$ step and seek for
the best-fit model by minimizing RMS of the residuals.
The light curves are also calculated for four companion masses of 
$M_{\rm MS}= 0.8$, 1.1, 1.5 and $2.0 M_\odot$.
Since we obtain similar light curves for all of these four cases,
we adopt $M_{\rm MS}= 1.5 M_\odot$ because the mass transfer 
is as high as $\sim 1 \times 10^{-7} M_\sun$ yr$^{-1}$ and this
indicates a thermally unstable mass transfer \citep{heu92},
which requires $M_{\rm MS}/M_{\rm WD} \gtrsim 1.1$ \citep{web85}.
A 1.5 $M_\sun$ MS expands to fill the Roche lobe of $\sim 1.7 ~R_\sun$
after a large part of hydrogen at the center is consumed 
\citep[e.g.,][]{hknu99}. 
\par
     In the best-fit model, 
a rather blue color index of $(B-V)_c \approx 0.14$ is obtained
as shown in Figure \ref{color_quiescence}.  
This suggests a large color excess of 
$E(B-V)= (B-V)_o - (B-V)_c \approx 0.86$
with the observed color of $(B-V)_o \sim 1.0$ 
\citep{men95}.
This color excess is very consistent with $E(B-V) = 0.85$ 
estimated by \citet{kis01}.

\placefigure{vmag_quiescence}
\placefigure{color_quiescence}

\section{THE 1917 OUTBURST}
     First we reproduce the light curve
of the 1917 outburst.  Its Tycho $B$ magnitude light curve 
has recently been reported by \citet{wil00db}, which is shown 
in Figure \ref{bmag_mmix_ciaql1917} together with our calculated
$B$ light curves.   Our numerical method for calculating light curves 
is fully described in \citet{hac01k}.
\par
     It has been established that the WD photosphere expands
to a giant size at the optical maximum in nova outbursts.  
Since the separation of
CI Aql is $a \sim 4~R_\sun$ \citep{gre96}, the WD envelope
engulfs the MS companion so that the early phase light curve 
is calculated only from the WD photosphere 
\citep[e.g.,][]{kat94h, kat99}. 
The accretion disk and the companion do not contribute to the $B$ light
until it decreases to $B\sim 13$ mag.
We determine the WD mass to be $M_{\rm WD}= 1.2 \pm 0.05 ~M_\sun$ 
from the early phase light curve 
as seen in Figure \ref{bmag_mmix_ciaql1917}.
\par
     Assuming $M_{\rm WD}= 1.2~M_\sun$, we estimate 
the envelope mass at the optical maximum to be
$\Delta M = 5.8 \times 10^{-6} M_\odot$ for the hydrogen
content $X=0.7$.  If the duration of quiescence before the 1917 outburst 
is almost the same as the quiescent phase between the 1917 and
2000 outburst, this envelope mass indicates a mass accretion rate of 
$\dot M_{\rm acc} \sim 0.7 \times 10^{-7} M_\odot$ yr$^{-1}$
if no WD matter has been dredged up.   In our wind model,
about 81\% ($4.7 \times 10^{-6} M_\odot$) of the envelope mass 
has been blown in the wind while the residual 19\% 
($1.1 \times 10^{-6} M_\odot$) has been left and added 
to the helium layer of the WD.  The net mass-increasing rate 
of the WD is $\dot M_{\rm He}= 0.13 \times 10^{-7} M_\odot$ yr$^{-1}$.
Thus, CI Aql may become an SN Ia if the donor is massive enough.

\placefigure{bmag_mmix_ciaql1917}

\section{THE 2000 OUTBURST}
     CI Aql bursted in April 2000.  
Its optical maximum ($V \sim 9$ mag) was reached on May 5, 2000
\citep[HJD 2451669.5,][]{kis01}.  
As shown in Figure \ref{vmag_irradmix_ciaql00},
the visual brightness quickly decreased to 13.5 mag 
in about 60 days.  Then, it stays at $V \sim 14$ mag, i.e.,
a plateau phase.  This mid-plateau phase is very similar 
to that of U Sco and can also be explained by the irradiation 
of the accretion disk \citep{hkkm00}.
Therefore, we have reproduced the light curve in the plateau phase
by assuming similar model parameters to those of U Sco.
Theoretical light curves in Figure \ref{vmag_irradmix_ciaql00}
are calculated for a pair of $1.2 ~M_\odot$ WD $+$ $1.5 ~M_\odot$ MS.
In the plateau phase of the light curve, i.e., when the WD photosphere
becomes much smaller than the binary size, 
the light curve is determined mainly 
by the irradiations of the ACDK and of the MS. 
The unheated surface temperatures are fixed 
to be $T_{\rm ph, MS}= 6300$ K for the MS 
and to be $T_{\rm ph, disk}= 5300$ K at the disk rim, which are
the same as the basic model in quiescence.
\par
     The luminosity of the accretion disk depends on 
both the thickness $\beta$ and the size $\alpha$.
We have examined a total of 160 cases for the set 
of $(\alpha, \beta)$, which is the product of 16 cases 
of $\alpha=$ 0.5---2.0 by 0.1 step
and 10 cases of $\beta=$ 0.05---0.50 by 0.05 step.
Here, we have adopted a set of $\alpha=1.4$ and $\beta=0.2$
during the wind phase and a set of $\alpha=0.7$ and $\beta=0.2$
after the wind stopped.
\par
     The decline of the early phase ($t \sim 0$---70 days) 
hardly depends on the hydrogen content $X$ of the WD envelope 
but the ends of the wind phase and of the hydrogen shell-burning phase
depend sensitively on the hydrogen content $X$ as shown 
in Figure \ref{vmag_irradmix_ciaql00}.  
\citet{mat00k} reported a sharp $\sim 1$ mag drop 
of $R$ magnitude in late November 2000 and Kiyota (VSNET archives)
also observed a $\sim 1.5$ mag drop of $I$ magnitude
as shown in Figure \ref{vmag_irradmix_ciaql00}.
If we attribute this drop to the end of wind phase,
the hydrogen content is somewhere between $X=0.35$ and
$X=0.5$, more closer to $X=0.5$.  We here adopt $X=0.5$.
In the case of $X=0.5$, the continuous wind stopped 
in early December 2000 about 220 days after maximum
and the steady hydrogen shell-burning will end 
in early August 2001 about 450 days after maximum
as shown in Figure \ref{vmag_irradmix_ciaql00}.

\placefigure{vmag_irradmix_ciaql00}

\section{SUPERSOFT X-RAY SOURCE PHASE}
     We here strongly encourage X-ray observations 
until the mid August 2001, that is, the end of steady hydrogen 
shell-burning.
No supersoft X-rays are observed in the early decline phase
because the photospheric radius of the WD is rather large 
and the photospheric temperature is relatively low.
After the recurrent nova enters the plateau phase, 
the photospheric temperature of the WD increases
to high enough to emit supersoft X-rays.  
During the massive wind phase, however, we do not expect
supersoft X-rays because they are self-absorbed by the wind itself. 
U Sco was observed as a luminous supersoft X-ray source
in the mid-plateau phase of the 1999 outburst just after
the massive wind stopped \citep{kah99, hkkm00}.
Our analysis suggests that the massive wind stopped in early 
December 2000 and that supersoft X-rays are now very likely detectable 
until August 2001.

\section{DISCUSSION}
     The calculated distance modulus of the 1917 outburst
is about $(m-M)_B= 14.5$.
Therefore, the distance to CI Aql is estimated to be 1.6 kpc for
the reddening of $E(B-V)= 0.85$ 
and the absorption of $A_B=3.5$ and $A_V=2.7$.
The apparent distance modulus of $(m-M)_V= 13.7$ from the 2000 outburst
as shown in Figure \ref{vmag_irradmix_ciaql00} is also consistent
with the distance of 1.6 kpc and $A_V=2.7$.
\par
     On the other hand, we obtain an apparent distance modulus 
of $(m-M)_V= 12.86$ in the quiescent phase as shown in Figure
\ref{vmag_quiescence}, which corresponds to the distance of 1.1 kpc.
This is not consistent with our estimation from the outburst light curves.
To examine this difference, we check other observations and found that 
three others indicate $V \approx 16.2$ in quiescence, which is 0.7 mag 
darker than $V \approx 15.5$ in Figure 1 of Mennickent \& Honeycutt 
\citep[see Table 2 of][]{men95, szk92}.
The $V$ magnitudes in their Figure 1 
were obtained with a fully automated telescope
and the errors in the zero point (formally 0.2 mag as in their text) 
could be larger than expected (private communication with Mennickent).
If we adopt $V = 16.2$ in quiescence, then the distance modulus
is $(m-M)_V= 13.56$ and the distance to CI Aql becomes $d= 1.5$ kpc,
being consistent with the distance estimate from the outbursts.
Thus, we determine the distance to CI Aql 
to be $d \sim 1.6$ kpc with $A_V= 2.7$.
\par
\begin{figure}
\plotone{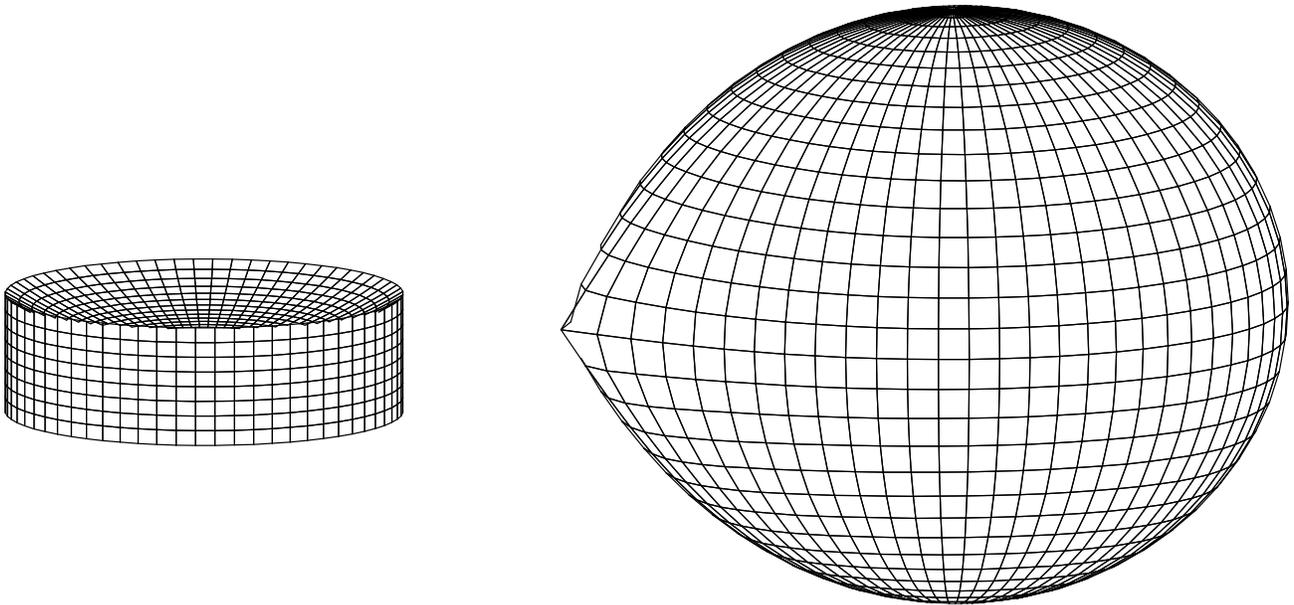}
\caption{
Configuration of our CI Aql model in quiescence.
The cool component ({\it right}) is a slightly evolved 
MS ($1.5 M_\odot$) filling up its inner critical Roche lobe.  
Only north and south polar areas of the cool component are 
irradiated by the hot component ($1.2~M_\odot$ WD, {\it left}).
The hot component itself and the central part of the ACDK
are not seen from the Earth because they are blocked 
by the flaring-up rim of the ACDK.  
Here the separation is $a= 4.25 R_\odot$; 
the effective radii of the inner critical Roche lobes are
$R_1^*= 1.53 R_\odot$, and $R_2^*= R_2= 1.69 R_\odot$, 
for the primary WD and the secondary MS, respectively.
\label{ciaql_quiescence_config}}
\end{figure}
\begin{figure}
\plotone{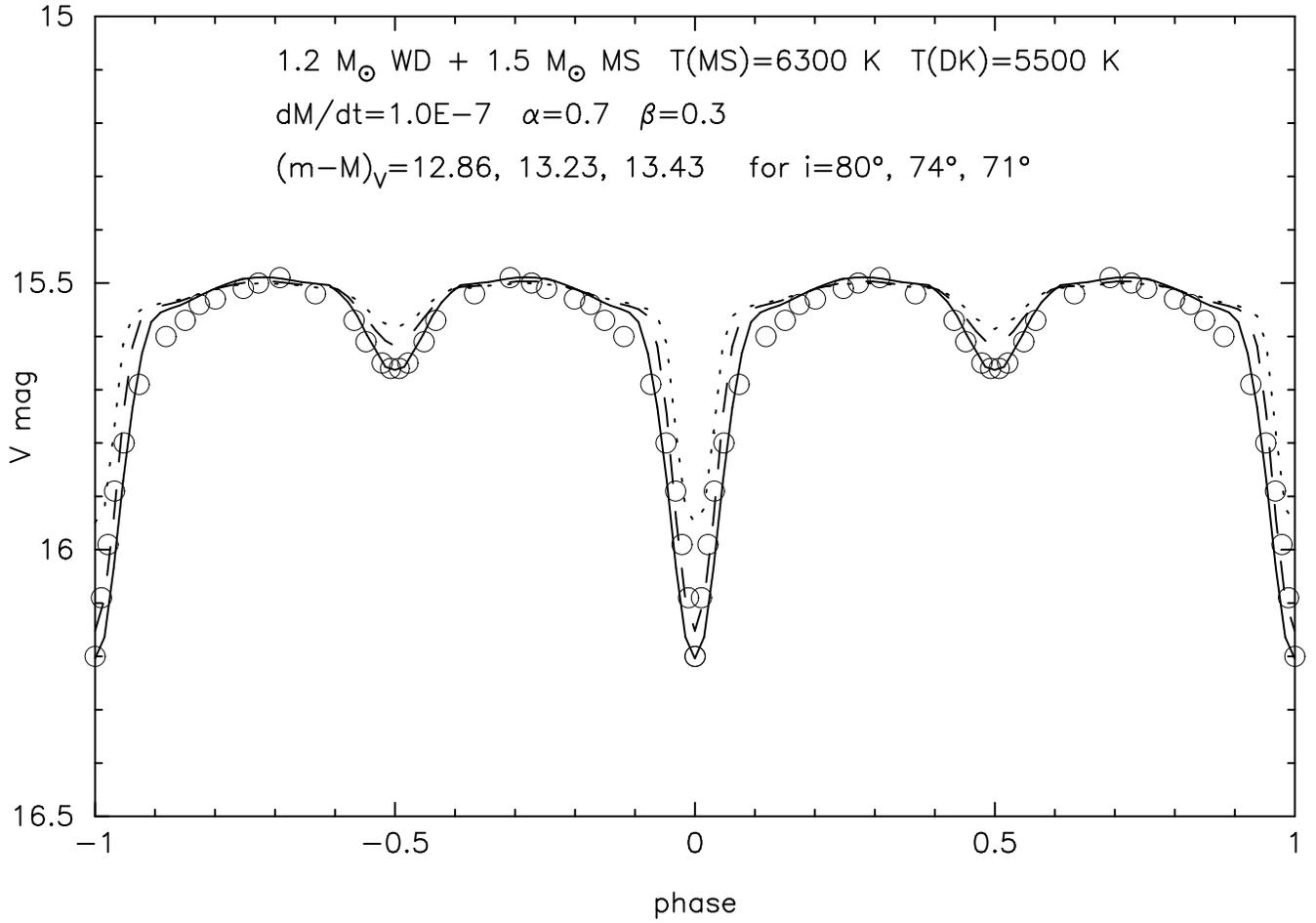}
\caption{
Calculated $V$ light curves are
plotted against the binary phase (two phases from $-1.0$ to $1.0$)
together with the smoothed observational points (open circles)
taken from \citet{men95}.
The model is a binary system of $1.2 M_\odot$ WD $+$ $1.5 M_\odot$ MS. 
The other parameters are printed in the figure. 
Light curves are plotted for different three inclination angles,
i.e., $i= 71\arcdeg$ (dotted), $74\arcdeg$ (dashed), 
and $80\arcdeg$ (solid).
\label{vmag_quiescence}}
\end{figure}
\begin{figure}
\plotone{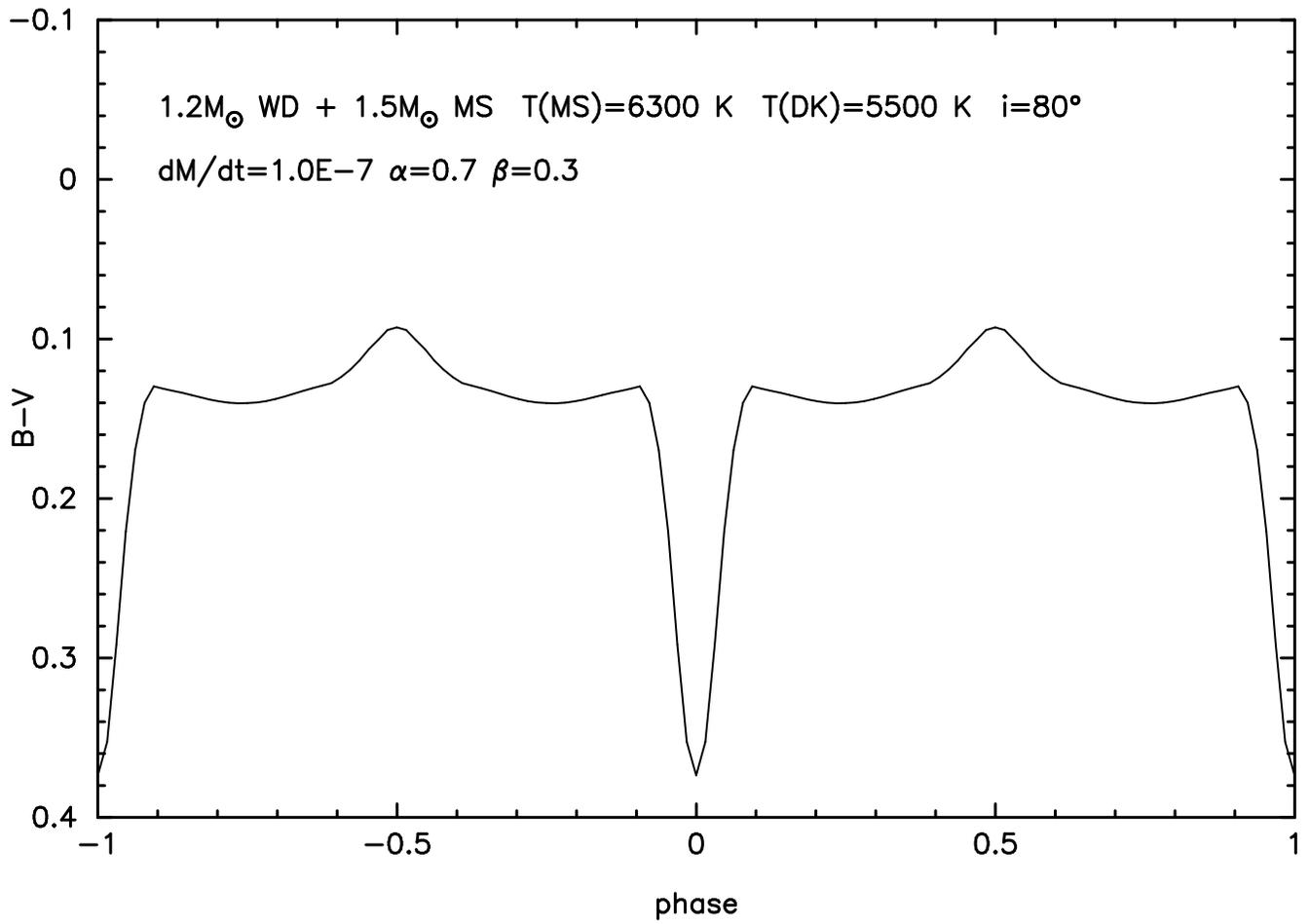}
\caption{
Calculated $B-V$ light curve against the binary phase
for the inclination angle of $i=80\arcdeg$.
\label{color_quiescence}}
\end{figure}
\begin{figure}
\plotone{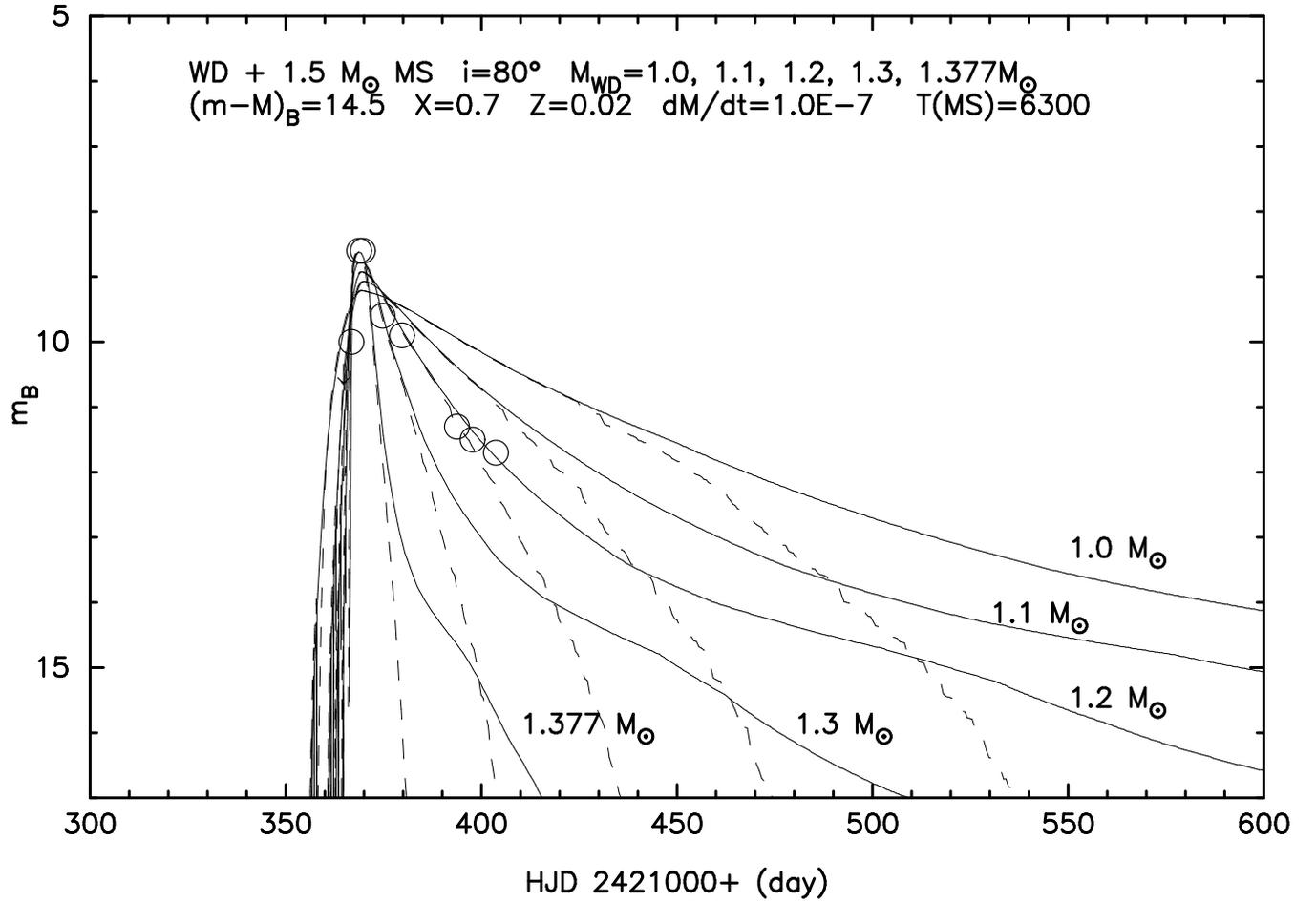}
\caption{
Calculated $B$ light curves are plotted against time (HJD 2,421,000+) 
together with the observational points of the 1917 outburst.
Open circles indicate observational points \citep[taken from][]{wil00db}.
The model consists of a bloated WD photosphere with no accretion
disk and a non-irradiated MS companion. 
The hydrogen content of the WD envelope is assumed to be $X=0.70$
for all models.  
The apparent distance modulus of $(m-M)_B= 14.5$ is assumed 
for all the WD masses.
Solid lines indicate the light curves connecting the $B$ light 
at the binary phase 0.5 (roughly the mean brightness)
while dashed lines correspond 
to those connecting the $B$ light at the binary phase 0.0
(mid-eclipse).
\label{bmag_mmix_ciaql1917}}
\end{figure}
\begin{figure}
\plotone{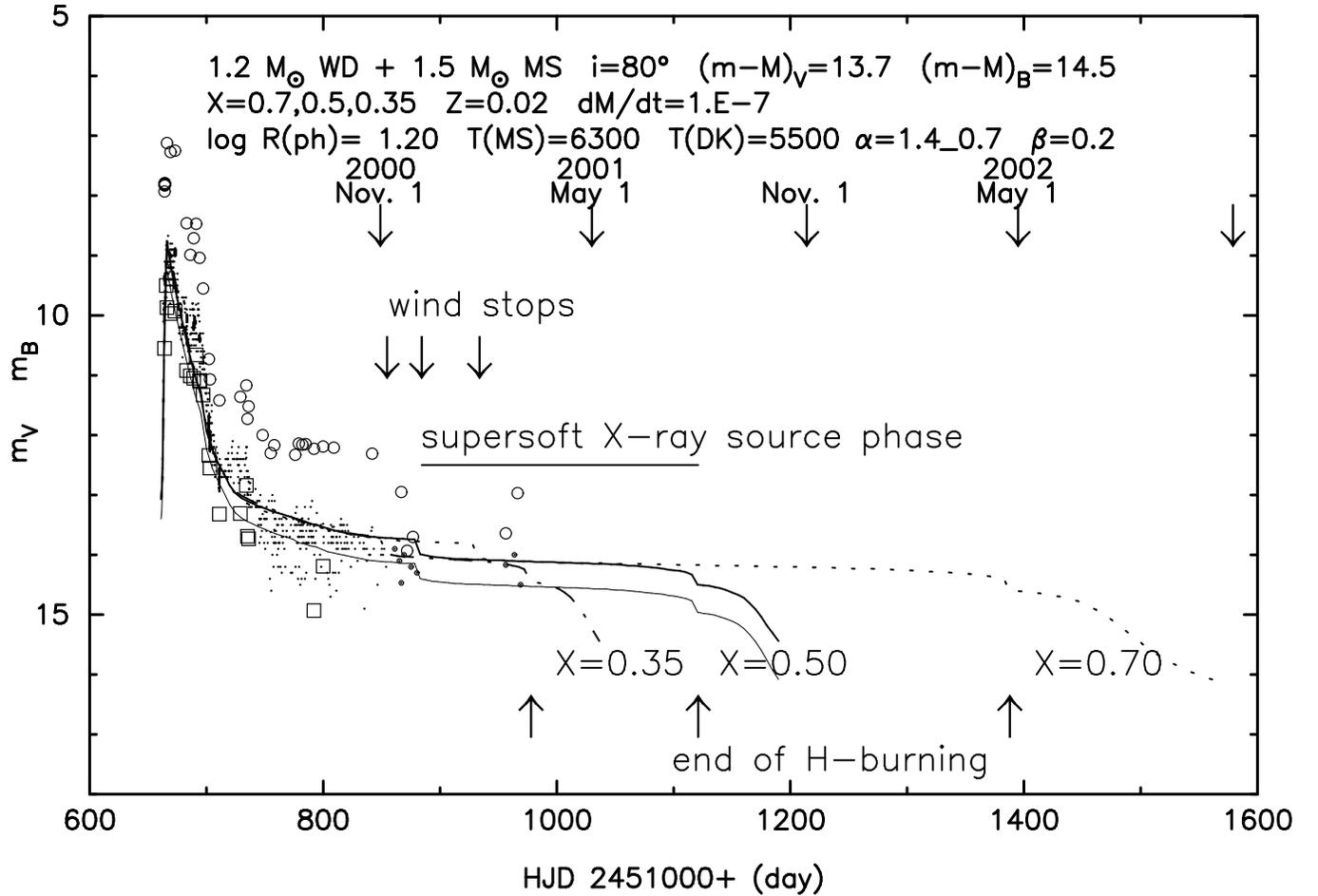}
\caption{
Calculated $V$ and $B$ light curves are plotted 
against time (HJD 2,451,000+) 
together with the observational points of the 2000 outburst.
Small dots indicate observational $V$ and visual magnitudes 
while open squares represent observational $B$ magnitudes
and open circles indicate observational $I$ magnitudes 
(all taken from VSNET archives).
Calculated $V$ light curves are plotted for 
three different hydrogen contents of the WD envelope,
$X=0.70$ (thick dotted), $X=0.50$ (thick solid), 
and $X=0.35$ (thick dot-dashed).
Thin solid line indicates $B$-magnitude of $X=0.50$ case.
Each light curve connects the brightness
at the binary phase 0.5 (roughly the mean brightness).
The apparent distance moduli of $(m-M)_V= 13.7$ and $(m-M)_B= 14.5$
are obtained by fitting.
\label{vmag_irradmix_ciaql00}}
\end{figure}

    The lacks of flickering and of Balmer emission lines in quiescence 
lead \citet{men95} to conclude that CI Aql is not interacting at least
at the present time.  However, the inner part of the accretion
disk cannot be seen from the Earth as easily understood from
Figure \ref{ciaql_quiescence_config}.  This is the reason for the lacks
of emission lines and of flickering even when the system includes
a mass-accreting WD.  This is essentially the same situation as in U Sco
\citep{sch90, joh92, sch95, hkkmn00}.



\acknowledgments
     We are very grateful to the VSNET members who observed CI Aql
and sent their valuable data to the VSNET.  We also thank R. Mennickent
for his comment on their photometry and the anonymous referee for 
many kind comments that help us to improve the manuscript.  
This research has been supported in part by the Grant-in-Aid for
Scientific Research (11640226) 
of the Japanese Ministry of Education, Science, 
Culture, and Sports.

\clearpage

\clearpage

\clearpage

\clearpage

\clearpage


\end{document}